\documentclass[amsmath,amssymb,a4paper,preprint,prl]{revtex4}
\usepackage{amssymb}
\usepackage{txfonts}
\usepackage{graphicx}
\usepackage{ulem}
\usepackage{color}

\begin{document}

\title{Phononic topological insulators with tunable pseudospin physics}

\author{Yizhou \surname{Liu}$^{1,2,3}$}
\author{Yong \surname{Xu}$^{1,2,3}$}
\email{yongxu@mail.tsinghua.edu.cn}
\author{Wenhui \surname{Duan}$^{1,2,4}$}

\affiliation{
$^1$State Key Laboratory of Low Dimensional Quantum Physics, Department of Physics, Tsinghua University, Beijing 100084, People's Republic of China \\
$^2$Collaborative Innovation Center of Quantum Matter, Beijing 100084, People's Republic of China \\
$^3$RIKEN Center for Emergent Matter Science (CEMS), Wako, Saitama 351-0198, Japan \\
$^4$Institute for Advanced Study, Tsinghua University, Beijing 100084, People's Republic of China}

\begin{abstract}
Efficient control of phonons is crucial to energy-information technology, but limited by the lacking of tunable degrees of freedom like charge or spin. Here we suggest to utilize crystalline symmetry-protected pseudospins as new quantum degrees of freedom to manipulate phonons. Remarkably, we reveal a duality between phonon pseudospins and electron spins by presenting Kramers-like degeneracy and pseudospin counterparts of spin-orbit coupling, which lays the foundation for ``pseudospin phononics''. Furthermore, we report two types of three-dimensional phononic topological insulators, which give topologically protected, gapless surface states with linear and quadratic band degeneracies, respectively. These topological surface states display unconventional phonon transport behaviors attributed to the unique pseudospin-momentum locking, which are useful for phononic circuits, transistors, antennas, etc. The emerging pseudospin physics offers new opportunities to develop future phononics.
\end{abstract}

\maketitle

Recently intensive research effort has been devoted to finding novel topological states of phonons, including the quantum anomalous Hall-like~\cite{prodan2009,zhang2010,wang2015njp,wang2015prl,yang2015,khanikaev2015,ni2015,peano2015,kariyado2015,fleury2016,susstrunk2016,huber2016,liu2017prb,liu2017nsr,abbaszadeh2017} and quantum spin Hall-like states~\cite{susstrunk2015,mousavi2015,wu2015,pal2016,he2016,liu2017prl}. These new quantum states of phonons are characterized by topologically protected, gapless boundary modes within the bulk gap, which are useful for various applications like high-efficiency phononic circuits/diodes and offer new paradigms for future phononics~\cite{huber2016,liu2017prb,liu2017nsr}. However, experimental realization of two-dimensional (2D) topological states is quite challenging. Specifically, the quantum anomalous Hall-like states require breaking time reversal symmetry of phonons, which remains experimentally illusive. The quantum spin Hall-like states rely on the pseudospin degeneracy protected by crystalline symmetries that typically get broken at the one-dimensional (1D) edges~\cite{liu2017prl}. In contrast, crystalline symmetries of three-dimensional (3D) systems can preserve simultaneously in the bulk and on the surface, enabling the topological protection. Importantly, most solid materials are crystalized in 3D lattices. Nevertheless, despite a few preliminary works on phononic topological semimetals~\cite{zhang2018,miao2018,li2018,xie2018}, 3D phononic topological insulators (TIs) have rarely been reported before, as far as we know. This is possibly because the spin (Kramers) degeneracy and spin-orbit coupling (SOC), which are essential to TIs , are missing for phonons. On the other hand, phonons are elementary excitations of lattice vibrations with zero charge and spin. The lacking of tunable degrees of freedom considerably limits their device applications. In this context, it is of critical importance to develop new quantum degrees of freedom for phonons. In light of the great success of spintronics, future research of phononics would be greatly enriched if one could establish any correlations between phonon pseudospins and electron spins \cite{li2012}.

In this Letter, we provided a guiding principle to design 3D phononic TIs as well as topological semimetals by utilizing crystalline symmetry-protected pseudospins characterized by Kramers-like degeneracy, quantized pseudoangular momenta and nonzero Berry curvature. Remarkably, we revealed pseudospin counterparts of the intrinsic and Rashba-Dresselhaus SOC, namely the pseudo-SOC of phonons, which builds a duality between phonon pseudospins and electron spins. The duality feature enables exploring the physics and applications of phonon pseudospins by borrowing ideas from spintronics, thus opening new opportunities for ``pseudospin phononics''.

\textit{Design principle of phononic TIs.}---An essential requirement of TIs is band degeneracies at no less than two high symmetry momenta (HSM) in the boundary Brillouin zone (BZ) \cite{fu2007prl,liu2014prb,dong2016}. The requirement is satisfied for electrons with spin degeneracies protected by time reversal symmetry. However, phonons do not have real spins, invoking different strategies for building phononic TIs. Naturally one could apply crystalline symmetries that are prevalent in solid materials to realize Kramers-like degeneracies. Such symmetries should also be preserved when projected onto the surface. However, for 2D spinless cases, no such kind of crystalline symmetry has higher than 1D irreducible representations at more than one HSM in the 1D edge BZ, implying that 2D phononic TIs protected by crystalline symmetries are forbidden \cite{liu2017prl}. The constraint is released for a variety of 3D lattices, where multiple band degenerate HSM can exist in the 2D surface BZ~\cite{fu2011,alexandradinata2014prl,liu2014prb,dong2016}. Thus the construction of 3D phononic TIs is feasible in principle.

\begin{figure}
\includegraphics[width=\linewidth]{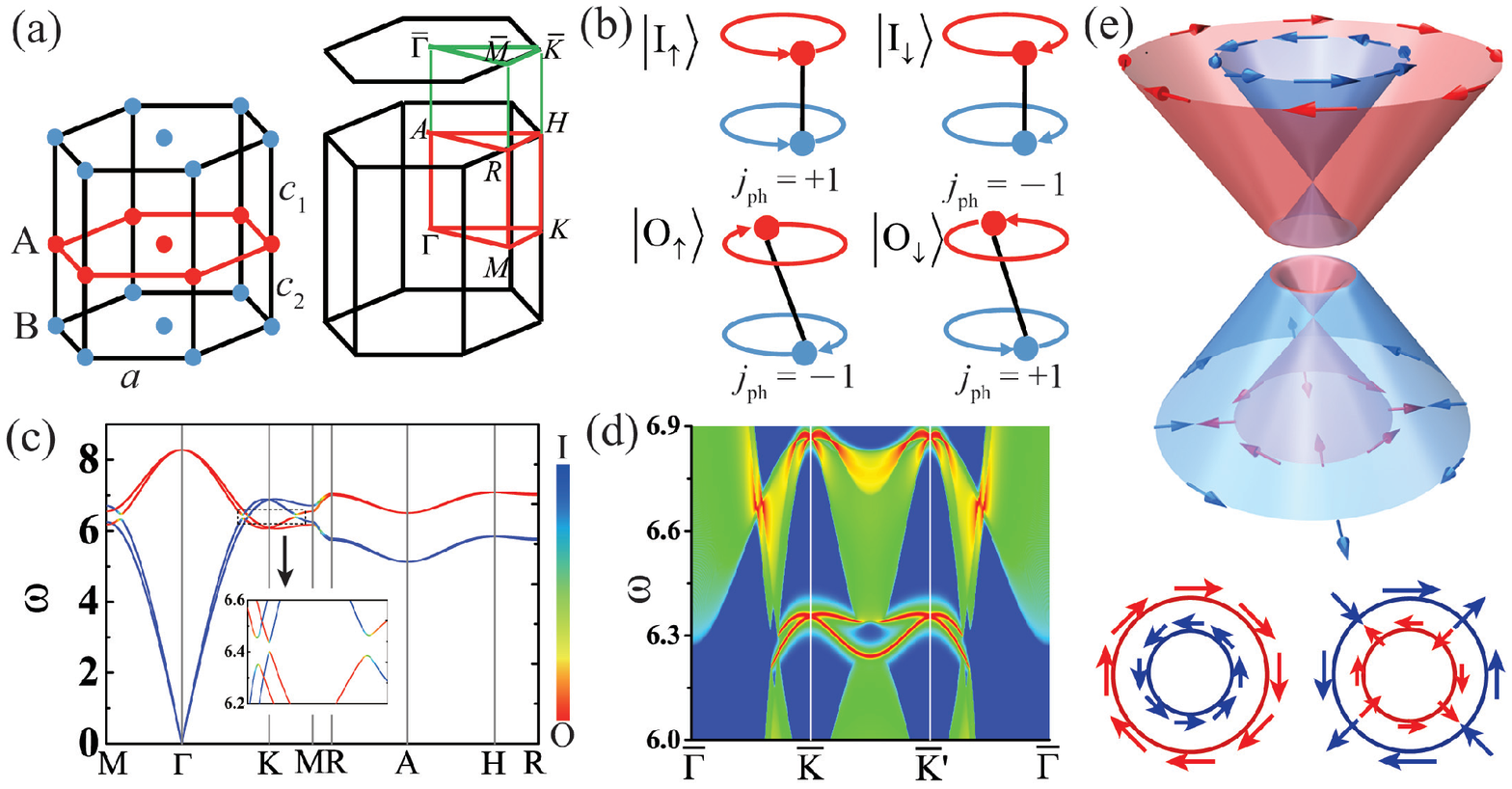}
\caption{\label{fig1}
(a) Atomic structure (left) and Brillouin zone (right) of a 3D triangular lattice with two sublattices A and B. Bond lengths are denoted by lattice parameters $a$, $c_1$ and $c_2$. (b) Schematic pseudospin states of in-phase ``I'' (out-of-phase ``O'') vibrational modes, whose pseudoangular momentum $j_{\text{ph}}$ is labelled. (c) Dispersion curves of a type-I phononic TI. Blue (red) color is used to denote the contribution of I (O) vibrational modes. (d) Local density of states (LDOS) of the (001) surface, where higher (lower) LDOS are colored red (blue). (e) Schematic phonon dispersion and pseudospin textures in the $k_z = 0$ plane near $K$. The bottom panel displays pseudospin textures of O (left) and I (right) vibrational modes from the top view.}
\end{figure}

The problem of phonons can be mapped to a tight-binding model of spinless electrons with fixed $p_{x,y,z}$ orbitals \cite{liu2017prb}. The symmetry representation of phonons is $\Gamma_{\text{phonon}} = \Gamma^{\text{equiv.}} \otimes \Gamma_{\text{vector}}$, where $\Gamma^{\text{equiv.}}$ is the equivalence representation of atomic sites and $\Gamma_{\text{vector}}$ is the representation of a 3D polar vector~\cite{dresselhaus2007}. Here we will not thoroughly discuss all possible 3D crystalline symmetries, but focus on $C_{nv}$ $(n = 3, 4, 6)$ symmetries that show interesting topological physics for electrons~\cite{fu2011,alexandradinata2014prl}. Take $C_{6v}$ lattices as an example. $\Gamma^{\text{equiv.}}$ is a $N$ dimensional representation, where $N$ is the number of atomic sites in a unit cell. When all the atomic sites are inequivalent, $\Gamma^{\text{equiv.}}$ is decomposed into $N$ 1D irreducible representations $A_1$. $N \ge 2$ is necessary to build a phononic TI that requires at least four bands. For simplicity, we choose $N = 2$ as displayed in Fig.~\ref{fig1}(a), which gives $\Gamma^{\text{equiv.}} = 2 A_1$. In the momentum space, the high-symmetry lines $\Gamma$-$A$ and $K$-$H$ in the bulk BZ are projected onto the HSM $\bar{\Gamma}$ and $\bar{K}$ in the surface BZ [Fig.~\ref{fig1}(a)], which have $C_{6v}$ and $C_{3v}$ symmetries, respectively. Under $C_{6(3)v}$ symmetry, $\Gamma_{\text{vector}} = A_1 \oplus E$, where $A_1$ and $E$ are 1D and 2D irreducible representations for basis functions of $p_z$ and $p_{x} \pm i p_{y}$, respectively. Therefore, $\Gamma_{\text{phonon}} = 2A_1 \oplus 2E$, implying that $p_{x} \pm i p_{y}$ ($p_z$) are doubly degenerate (nondegenerate) along the high-symmetry lines. Note that frequencies of in-plane vibrations ($p_{x} \pm i p_{y}$) are typically higher than those of out-of-plane ones ($p_z$), leading to weak hybridization between the doublet and singlet states. We thus focus on in-plane vibrations~\cite{suppl}, for which a $\mathbb{Z}_2$ classification of band topology is permitted~\cite{fu2011}.

\textit{Pseudospin- and topology-related physics.}---Considering that the double degeneracy of $p_x \pm ip_y$ resembles the Kramers degeneracy of spins, we introduce a pseudospin index to label the Kramers-like states. There exist two types of phonon modes, including in-phase (I) and out-of-phase (O) vibrations between the two atomic sites A and B. Their pseudospin states [Fig. \ref{fig1}(b)] are defined in coordinates of $(x_A, y_A, x_B, y_B)^T$ as
\begin{equation}
\begin{split}
|\text{I}_{\uparrow}\rangle &= (\varepsilon_+ \sin\theta_{\textbf{k}}, \varepsilon_+ \cos\theta_{\textbf{k}})^T,  \ |\text{O}_{\uparrow}\rangle = (\varepsilon_- \cos\theta_{\textbf{k}}, -\varepsilon_- \sin\theta_{\textbf{k}})^T, \\
|\text{I}_{\downarrow}\rangle &= i(\varepsilon_- \sin\theta_{\textbf{k}}, \varepsilon_- \cos\theta_{\textbf{k}})^T,  |\text{O}_{\downarrow}\rangle = i(-\varepsilon_+ \cos\theta_{\textbf{k}}, \varepsilon_+ \sin\theta_{\textbf{k}})^T, \nonumber
\end{split}
\end{equation}
where $\varepsilon_\pm = (1,\pm i)/\sqrt{2}$, and the vibrational magnitude of each atomic site is determined by $\theta_{\textbf{k}} \in [0, \pi/2]$ that is $\textbf{k}$-dependent. They are orthonormal and form a complete basis of in-plane vibrations.

The pseudospin states are featured by well-defined, quantized pseudoangular momenta about $z$ axis for $\textbf{k}$ along the high-symmetry lines due to the $C_{6(3)}$ rotational symmetry. The pseudoangular momentum operator is expressed as $J_{\text{ph}} = \sigma_z s_z$~\cite{suppl}, where $\sigma_z=\pm1$ and $s_z=\pm1$ refer to pseudospin up (down) and I (O) vibrational modes, respectively. The phonon pseudoangular momentum $j_{\text{ph}}$ is composed of two parts, including a local part determined by the on-site orbital and a nonlocal part contributed by the inter-site Bloch phase change~\cite{zhang2015}. Herein the nonlocal part is zero and $j_{\text{ph}}$ is fully determined by the local part, giving $j_{\text{ph}} = 1$ for $|\text{I}_{\uparrow}\rangle$ and $|\text{O}_{\downarrow}\rangle$ and $j_{\text{ph}} = -1$ for $|\text{I}_{\downarrow}\rangle$ and $|\text{O}_{\uparrow}\rangle$, as depicted in Fig. \ref{fig1}(b).

Next we derived the effective Hamiltonian around HSM by symmetry analysis using the basis set \{$|\text{I}_{\uparrow}\rangle, |\text{O}_{\uparrow}\rangle, |\text{I}_{\downarrow}\rangle, |\text{O}_{\downarrow}\rangle$\}. Derivation details were presented in Supplemental Material~\cite{suppl}. The HSM are classified into type-I ($K/K'$ and $H/H'$) and type-II ($\Gamma$ and $A$), which have $C_{3v}$ and $C_{6v}$ symmetries, respectively, and will be discussed separately. Hereafter the wave vector $\mathbf{k}$ is referenced to the HSM. The effective Hamiltonian near the type-I HSM is written as $H = H_0 + H_I + H_{RD}$, where $H_0 = \text{diag}(M_{\mathbf{k}}, -M_{\mathbf{k}}, M_{\mathbf{k}}, -M_{\mathbf{k}})$,
\begin{equation}\label{Heff1}
\begin{split}
H_I = &\left(
\begin{array}{cccc}
0                & \Delta_1k_+       & 0              & \Delta_2k_z  \\
\Delta_1k_-      & 0                 & \Delta_2k_z    & 0 \\
0                & \Delta_2k_z       & 0              & -\Delta_1k_-  \\
\Delta_2k_z      & 0                 & -\Delta_1k_+   & 0
\end{array}
\right), \\
H_{RD} = &\left(
\begin{array}{cccc}
0         & 0          & -iC_1k_+   & 0 \\
0         & 0          & 0          & iC_2k_- \\
iC_1k_-   & 0          & 0          & 0 \\
0         & -iC_2k_+   & 0          & 0
\end{array}
\right),
\end{split}
\end{equation}
$M_{\mathbf{k}} = M - B_1(k^2_x+k^2_y) - B_2k^2_z$, $k_\pm=k_x\pm ik_y$. The curvature parameters $B_1$ and $B_2$ typically have the same sign.

Remarkably, this effective Hamiltonian resembles the 3D Bernevig-Hughes-Zhang (BHZ) model with broken inversion symmetry for electrons. $H_0 + H_I$ is exactly the same as the typical 3D BHZ model~\cite{zhang2009,qi2010,qi2011}, where $H_I$ is attributed to the intrinsic SOC. For electrons, $H_{RD}$ arises in conditions of broken inversion symmetry, which includes $\pm iC_2 k_\mp$ and $\mp iC_1 k_\pm$ terms, corresponding to the Rashba $(k_y \sigma _x  - k_x \sigma _y )$ and Dresselhaus $(k_y \sigma _x  + k_x \sigma _y)$ SOC, respectively. In analogy, $H_I$ and $H_{RD}$ are called the intrinsic and Rashba-Dresselhaus pseudo-SOC for phonon pseudospins. In this sense, the problems of phonons and electrons are described by essentially the same Hamiltonian, though their underlying physics is distinctly different. This implies a ``duality'' between phonon pseudospins and electron spins. Therefore, the topological and quantum physics revealed for electron spins is applicable for phonon pseudospins, and vice versa. This key result could lay the foundation for an emerging field of ``pseudospin phononics''.

Then we will discuss phononic topological properties by the effective Hamiltonian. When excluding $H_{RD}$ and selecting $k_z = 0$, $H$ reduces to the 2D BHZ model, which gives quantized pseudospin-resolved Chern numbers $\mathcal{C}_{{\uparrow}({\downarrow})}$ \cite{kane2005a}. As the type-I HSM exist in pairs, $\mathcal{C}_{{\uparrow}({\downarrow})} = \pm 2$ when a band inversion occurs (i.e. $M/B_1>0$). The sum of $\mathcal{C}^{s} = (\mathcal{C}_{\uparrow} - \mathcal{C}_{\downarrow})/4$ contributed by all HSM mod 2 gives a topological invariant $\mathbb{Z}_2$. The 3D phononic TI phase is characterized by $\mathbb{Z}_2 = 1$. The inclusion of $H_{RD}$ introduces intraband coupling between opposite pseudospins, which removes the  pseudospin degeneracy except at the HSM. Then $\mathcal{C}_{{\uparrow}({\downarrow})}$ gets ill defined, but the $\mathbb{Z}_2$ topological classification remains valid~\cite{fu2011}. The $\mathbb{Z}_2$ topological invariant will not be affected by $H_{RD}$ as far as the bulk band gap keeps open when adiabatically turning on $H_{RD}$.

To demonstrate the nontrivial topological states, we explicitly studied lattice vibrations in a $C_{6v}$ lattice [Fig.~\ref{fig1}(a)]. The interatomic interactions between the nearest and next-nearest neighbors were described by longitudinal and transverse force constants, as done previously~\cite{zhang2015}. The out-of-plane vibrations typically have minor influence on topological properties of in-plane vibrations~\cite{suppl}, which are thus neglected for simplicity. We systematically searched the whole space of interatomic coupling parameters, and found that the required band inversions can be obtained by a wide range of coupling parameters. Details of calculation methods and parameters related to the following discussions were described in Supplemental Material~\cite{suppl}.

Figure~\ref{fig1}(c) presents phonon dispersion curves with a band inversion between I and O vibrational modes at $K$. This band inversion leads to a nontrivial band topology $\mathbb{Z}_2 = 1$, as confirmed by our calculations of hybrid Wannier centers~\cite{taherinejad2014,alexandradinata2014prb} that display partner switching between Kramers-like pairs~\cite{suppl}. Moreover, there is a frequency gap between the two kinds of bands. The system is thus a 3D phononic TI. A hallmark of phononic TIs is the existence of gapless surface states within the bulk gap, which is topologically protected when the corresponding symmetry is preserved. On the (001) surface where the $C_{6v}$ symmetry preserves, we indeed observed a single pair of gapless Dirac-cone-shaped surface bands located near $\bar{K}$ and $\bar{K'}$ [Fig.~\ref{fig1}(d)], as warranted by the bulk-boundary correspondence~\cite{hasan2010}.

Figure~\ref{fig1}(e) displays schematic pseudospin textures of bulk bands in the $k_z = 0$ plane. There would be no net pseudospin polarization if excluding the Rashba-Dresselhause pseudo-SOC interaction $H_{RD}$. Interestingly, when including $H_{RD}$, Rashba-like and Dresshause-like pseudospin textures evolve in the O and I bands, respectively. We further considered the topological surface states (TSSs) near the Dirac point, which are described by the effective Hamiltonian (referenced to the Dirac frequency) $H_{\text{surf}} = v_D(k_x\sigma_x\tau_z + k_y\sigma_y)$, where $\tau_z=\pm1$ refers to $K$ ($K^\prime$) valley; $v_D$ is the group velocity at the Dirac point. Noticeably, $H_{\text{surf}}$ of each valley has the same form as for TSSs of electrons~\cite{zhang2009}, whose pseudospin textures are schematically displayed in Fig. \ref{fig3}(a). By adiabatically varying $\mathbf{k}$ along the loop enclosing $K$ ($K'$), the pseudospin vectors wind $\pm1$ times, giving quantized Berry phases of $\pm\pi$. The similarity between phonon pseudospins and electron spins is thus well demonstrated for both bulk and surface bands.

\textit{Type-II phononic TIs.}---Type-II phononic TIs are characterized by the existence of band inversions at type-II HSM ($\Gamma$ or $A$). Importantly, due to the $C_6$ rotation symmetry, all the linear terms of $k_\pm$ are forbidden in the effective Hamiltonian near type-II HSM, which is expressed as~\cite{suppl}:
\begin{equation}\label{Heff2}
H'= H_0 + H'_I + H'_{RD} = \left(
\begin{array}{cccc}
M_{\mathbf{k}} & \delta_1k^2_-      & -ic_1k^2_-     & \delta_2k_z \\
\delta_1k^2_+  & -M_{\mathbf{k}}    & \delta_2k_z    & ic_2k^2_+ \\
ic_1k^2_+      & \delta_2k_z        & M_{\mathbf{k}} & -\delta_1k^2_+ \\
\delta_2k_z    & -ic_2k^2_-         & -\delta_1k^2_- & -M_{\mathbf{k}}
\end{array}
\right).
\end{equation}
A new kind of intrinsic and Rashba-Dresselhaus pseudo-SOC ($H'_I$ and $H'_{RD}$) is thus introduced.

\begin{figure}
\includegraphics[width=\linewidth]{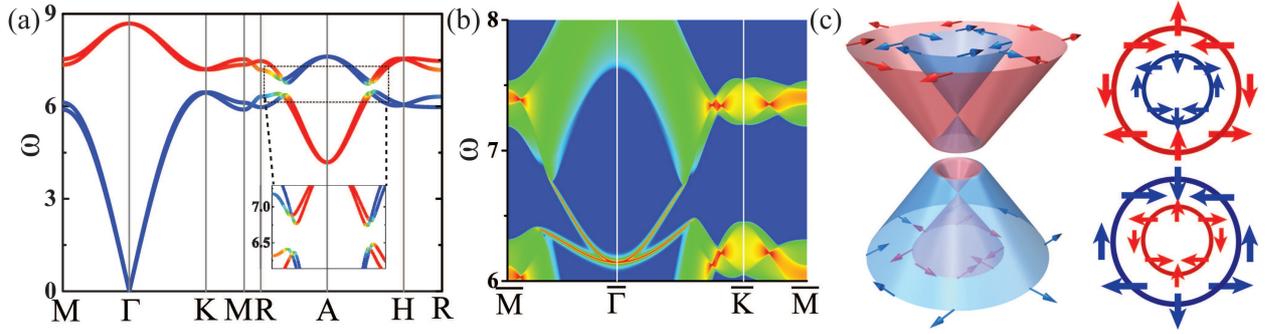}
\caption{\label{fig2}
(a) Dispersion curves of a type-II phononic TI. Blue (red) color is used to denote the contribution of I (O) vibrational modes. (b) LDOS of the (001) surface, where higher (lower) LDOS are colored red (blue). (c) Schematic phonon dispersion and pseudospin textures in the $k_z=0$ plane near $A$. The right panel displays pseudospin textures of O (top) and I (bottom) vibrational modes from the top view.}
\end{figure}

\begin{figure}
\includegraphics[width=\linewidth]{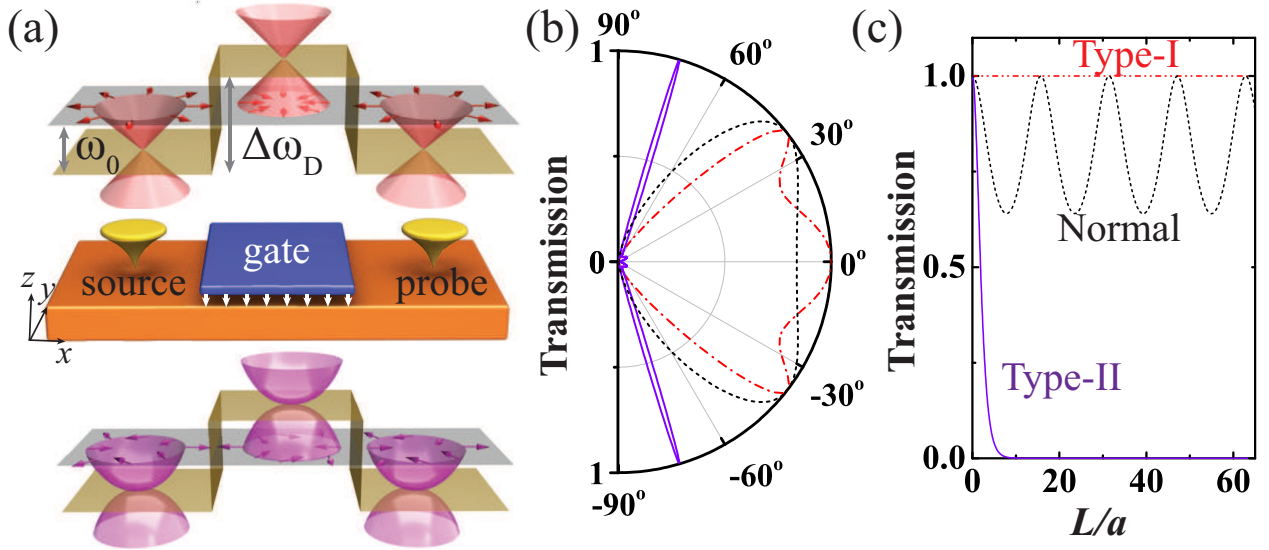}
\caption{\label{fig3}
(a) Schematic phonon tunnelling junction, where surface lattice vibrations of frequency $\omega_0$ are excited (detected) by a point-like source (probe) and a tunneling barrier $\Delta \omega_D$ is applied by a piezoelectric gate. Schematic phonon dispersion and pseudospin textures of type-I (type-II) TSSs are displayed in the top (bottom) panel. (b) Phonon transmission as a function of incident angle $\theta$ for type-I TSSs (red dash-dotted), type-II TSSs (purple solid) and normal surface states (black dashed), for which we used barrier widths of $L =$ $132a$, $29.5a$ and $132 a$, respectively. $a$ is the length parameter depicted in Fig. 1(a). $\theta=0$ is defined along the $x$-axis. (c) Phonon transmission as a function of $L$ for $\theta=0$. $\omega_0 = 0.1$ and $\Delta \omega_D = 0.3$ were used.}
\end{figure}

Figure~\ref{fig2}(a) presents dispersion curves of a type-II phononic TI, which is characterized by a band inversion at $A$, a finite frequency gap and a nontrivial band topology $\mathbb{Z}_2 = 1$ verified by the calculations of hybrid Wannier centers~\cite{suppl}. We also calculated the surface states of the (001) termination [Fig. \ref{fig2}(b)], which shows a quadratic band crossing at $\bar{\Gamma}$, in contrast to a pair of linear band crossings at type-I HSM. The type-II TSSs are described by the effective Hamiltonian (referenced to the degenerate frequency) $H'_{\text{surf}} =  D [(k^2_x - k^2_y)\sigma_x - 2k_xk_y\sigma_y]$, where $D$ is a coefficient. Pseudospin textures of bulk bands [Fig. \ref{fig2}(c)] are neither typical Rashba-like nor typical Dresselhaus-like, but display new pseudospin-momentum locked features. Pseudospin textures of type-II TSSs are significantly different from those of type-I TSSs [Fig. \ref{fig3}(a)], which are characterized by winding numbers of $\pm2$ and quantized Berry phases of $\pm2\pi$.

\textit{Pseudospin-momentum locked phonon transport}---One prominent feature caused by the pseudo-SOC $H_{RD}$ or $H'_{RD}$ is that the pseudospin and momentum of phonons are locked, which plays an important role in determining transport properties. Moreover, type-I and type-II bulk/surface bands are featured by different kinds of pseudospin-momentum locking, leading to distinct transport behaviors. To demonstrate the concept, we simulated phonon transport of TSSs in tunnelling junctions \cite{suppl}, where a tunneling barrier is introduced into the gate region as schematically displayed in Fig. \ref{fig3}(a). The height of tunneling barrier, in principle, could be controlled by using a piezoelectric gate that changes the on-site potential, strain, or interatomic coupling of surface atoms \cite{li2012}. Figure~\ref{fig3}(b) shows phonon transmissions as a function of incident angle $\theta$ for type-I and type-II TSSs. Normal surface states with a linear band dispersion and no pseudospin-momentum locking were also studied for comparison. Noticeably, there exists resonant phonon transport (transmission equal to 1) along some particular directions, which satisfy the constructive interference condition of Fabry-P\'erot resonances, $q_x L=n\pi$, where $q_x$ is the $x$-component of wavevector in the gate region (Supplemental Fig. S3), $L$ is the barrier width, and $n$ is an integer number. Moreover, phonon transmission of normal surface states oscillates with varying $L$, which is also expected by the Fabry-P\'erot physics.

However, substantially different results are found for type-I and type-II TSSs. Take $\theta=0$ for example. Transport of type-I TSSs keeps ballistic, while phonon transmission of type-II TSSs decays exponentially with increasing $L$ [Fig.~\ref{fig3}(c)]. These unconventional transport behaviors are insensitive to material details, indicating a topological origin. Importantly, these physical effects can be well understood by pseudospin physics. Specifically, for type-I (type-II) TSSs, the forward-moving phonon modes have the same (opposite) pseudospins between the upper and lower Dirac cones, which correspond to transport channels of the source/probe and gate, respectively [Fig.~\ref{fig3}(a)]. Because of the perfectly matched (mismatched) pseudospins, phonons are able to transport across the barrier with no (full) backscattering. These quantum transport phenomena are inherently related to the quantized Berry phase $\pi$ ($2\pi$) of type-I (type-II) TSSs, which induces destructive (constructive) interferences between incident and backscattered waves.

Our findings suggest some potential applications of TSSs. For instance, the excellent transport ability of type-I TSSs can be utilized for low-dissipation phononic devices. The strongly angle-dependent transmission of type-II TSSs, which are confined to transport along some specific directions, can be used to design directional phononic antenna. Moreover, type-II TSSs are promising for building efficient phononic transistors, because their phonon conduction in the tunnelling junction can be switched off (on) by a finite (zero) barrier.

\begin{figure}
\includegraphics[width=\linewidth]{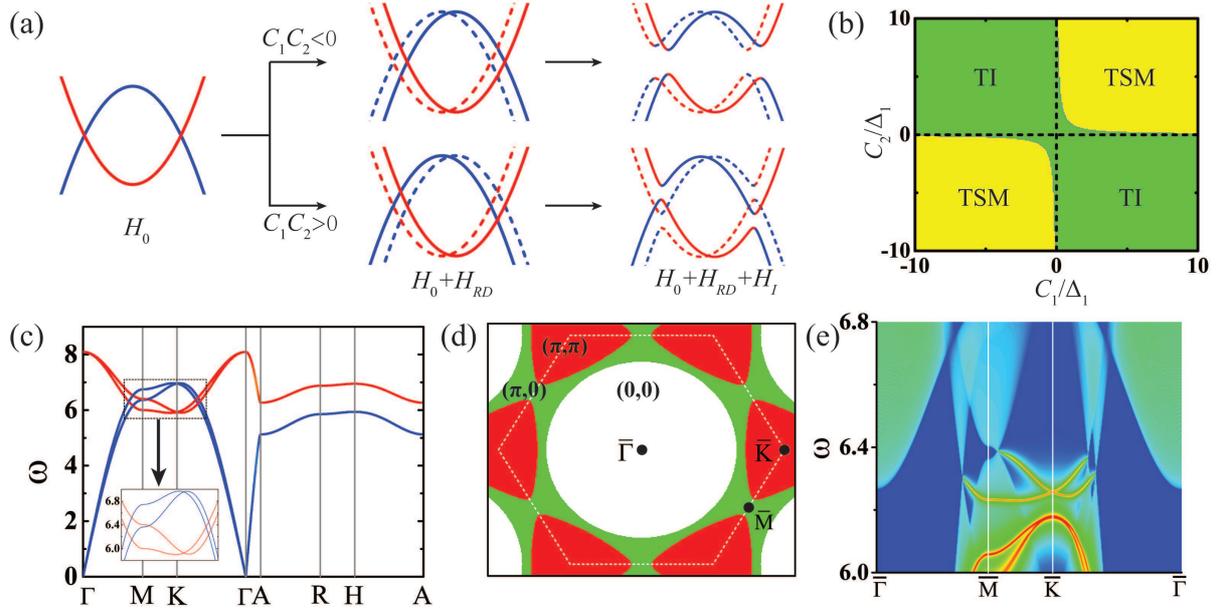}
\caption{\label{fig4}
(a) Schematic evolution of I (blue) and O (red) vibrational bands near $K$ under the influence of intrinsic pseudo-SOC $H_I$ and Rashba-Dresselhause pseudo-SOC $H_{RD}$. The $H_{RD}$-induced pseudospin splittings of I and O vibrational bands have same (opposite) signs when $C_1 C_2 < 0$ ($C_1 C_2 > 0$), which can result in phononic TIs or topological semimetals (TSMs) as summarized in the topological phase diagram (b). (c) Dispersion curves of a phononic topological nodal-ring semimetal. Blue (red) color is used to denote the contribution of I (O) vibrational modes. (d) Zak phases along the (001) direction of the lowest two bands for the 2D surface BZ. (e) LDOS of the (001) surface, where higher (lower) LDOS are colored red (blue).}
\end{figure}

\textit{Phononic topological semimetals.}---In addition to phononic TIs, topological band inversions can introduce other novel topological phases, including topological nodal-ring semimetals and topological Weyl semimetals, which are collectively called phononic topological semimetals. The essential physics is illustrated in Fig.~\ref{fig4}(a). Let us start from $H = H_0$ and take two pairs of bands inverted at $K$ for example. Generally, the pseudospin degeneracy is split by $H_{RD}$, and a full band gap is induced by $H_{I}$, leading to phononic TIs. However, phononic topological semimetals would emerge if the opening of band gap was forbidden by symmetry. This is possible here, providing that band splittings of $H_{RD}$  have opposites signs between I and O vibrational modes (i.e. $C_1C_2>0$). Then, when in the presence of mirror symmetry $M_z$ (with the same atoms at A and B sites), the crossing rings in the $k_z = 0$ plane are protected to be gapless, introducing topological nodal-ring semimetals. In contrast, under broken $M_z$, the nodal rings are gapped out except for some gapless points that are protected to exist in the $\Gamma$-$K$ line by mirror symmetry $M_x$. These gapless points are Weyl points, and the resulting phase is topological Weyl semimetal. A unified description of these topological phases is provided by the phase diagram presented in Fig.~\ref{fig4}(b).

Figure~\ref{fig4}(c) presents dispersion curves of a phononic topological nodal-ring semimetal protected by $M_z$, for which equivalent A and B sites were selected \cite{suppl}. The gapless feature in the $k_z = 0$ plane is consistent with the symmetry analysis.  Moreover, we calculated the Zak phase $\theta_{\text{Zak}}$ along the (001) direction for the 2D surface BZ, where $\theta_{\text{Zak}}$ has quantized values of 0 or $\pi$ [Fig. \ref{fig4}(d)]. The nodal rings appear at boundaries between regions of $\theta_{\text{Zak}} = 0$ and $\theta_{\text{Zak}} = \pi$, implying their topological nature. Furthermore, phononic drumhead surface states were observed by our surface-state calculations [Fig. \ref{fig4}(e)], which is a hallmark feature of topological nodal-ring semimetals. Their evolution with varying phonon frequencies are displayed in Supplemental Fig. S5.

\textit{Outlook.}---Our work sheds lights on future study of topological phononics. A few promising research directions are opened: (i) To search for new symmetry-protected phononic topological states. In additional to $C_{nv}$, there are many other symmetries, like (magnetic) space group symmetry, time reversal symmetry, particle-hole symmetry, etc, and their combinations, which could result in rich topological phases \cite{liu2014prb,fang2015}. (ii) To explore novel physical effects by breaking symmetry locally or globally. For instance, the quantum anomalous Hall-like states would emerge if time reversal symmetry breaking effects were introduced to 2D TSSs of phononic TIs. (iii) To investigate unconventional electron-phonon coupling and superconductivity caused by the pseudospin- and topology-related physics of phonons. Many material systems could simultaneously host nontrivial topological states of electrons and phonons. Thus, TSSs of electrons and phonons are able to coexist on the material surfaces. Their interactions might be exotic, for instance, due to the (pseudo-)spin-momentum locking. (iv) To find realistic candidate materials for experiment and application. Very few realistic materials of chiral phonons~\cite{zhang2015,liu2017prl,zhu2018} and phononic topological semimetals~\cite{xie2018,zhang2018,li2018,miao2018} have been reported, plenty of unknown candidate materials of various phononic topological phases are awaiting to be discovered. It is helpful to develop machine learning methods for high throughput discovery of phononic topological materials.

\begin{acknowledgments}
This work was supported by the Basic Science Center Project of NSFC (Grant No. 51788104), the Ministry of Science and Technology of China (Grants No. 2016YFA0301001, No. 2018YFA0307100 and No. 2018YFA0305603), the National Natural Science Foundation of China (Grants No. 11674188 and No. 11334006), and the Beijing Advanced Innovation Center for Future Chip (ICFC). Y.X. acknowledges support from the National Thousand-Young-Talents Program and Tsinghua University Initiative Scientific Research Program.
\end{acknowledgments}


\end{document}